\documentclass[aps,prd,preprintnumbers,10pt,superscriptaddress,twocolumn,nofootinbib]{revtex4-2}

\usepackage[dvips]{graphicx}

\usepackage{braket}

\usepackage{siunitx}
\usepackage{array,amsmath,amsthm,feynmf,mathtools}
\usepackage{multirow}
\usepackage{mathtools}
\usepackage{epsfig}
\usepackage{graphicx}
\usepackage{xcolor}
\usepackage{slashed}
\usepackage{amssymb}
\usepackage{bm}
\usepackage{cancel}
\usepackage[normalem]{ulem}
\usepackage[export]{adjustbox}
\usepackage{enumitem}
\usepackage[colorlinks=true,linkcolor=blue,citecolor=blue]{hyperref}
\usepackage{float}

\usepackage{hhline}
\usepackage[column=O]{cellspace}
\renewcommand{\arraystretch}{1.2}

\newcommand{\lagr}{\mathcal{L}}

\newcommand{\ord}[1]{\mathcal{O}(#1)}

\newcommand{\SM}{SM }
\newcommand{\CW}{clockwork }

\begin{document}
\title{A Supersymmetric Flavor Clockwork}
\author{Wolfgang Altmannshofer}
\email{waltmann@ucsc.edu}
\affiliation{Department of Physics and Santa Cruz Institute for Particle Physics\\ University of California, Santa Cruz, CA 95064, USA}

\author{Sri Aditya Gadam}
\email{sgadam@ucsc.edu}
\affiliation{Department of Physics and Santa Cruz Institute for Particle Physics\\ University of California, Santa Cruz, CA 95064, USA}

\begin{abstract}
Clockwork models can explain the flavor hierarchies in the Standard Model quark and lepton spectrum.
We construct supersymmetric versions of such flavor clockwork models. The zero modes of the clockwork are identified with the fermions and sfermions of the Minimal Supersymmetric Standard Model. In addition to generating a hierarchical fermion spectrum, the clockwork also predicts a specific flavor structure for the soft SUSY breaking sfermion masses. We find sizeable flavor mixing among first and second generation squarks. Constraints from Kaon oscillations require the masses of either squarks or gluinos to be above a scale of $\sim 3$ PeV.
\end{abstract}

\maketitle


\section{Introduction}

Among the many motivations for physics beyond the Standard Model (SM), there is the hierarchical structure of the quark and lepton spectrum that remains unexplained in the SM.
In fact, the SM Yukawa couplings span several orders of magnitude, from an $\mathcal O(1)$ top quark Yukawa coupling to an electron Yukawa coupling of a few $\times 10^{-6}$. Numerous mechanisms have been invoked in the literature to explain such a hierarchy, including frameworks with spontaneously broken $U(1)$ flavor symmetries~\cite{Froggatt:1978nt, Leurer:1992wg, Leurer:1993gy, Grossman:1995hk}, localization of fermions in extra dimensions~\cite{ArkaniHamed:1999dc, Grossman:1999ra, Gherghetta:2000qt, Kaplan:2001ga}, models with large anomalous dimensions \cite{Nelson:2000sn}, and models that explain the hierarchies with loop factors~\cite{Weinberg:1972ws, Balakrishna:1988ks} (see~\cite{Altmannshofer:2014qha, Bauer:2015fxa, Knapen:2015hia, Alonso:2016onw, vonGersdorff:2017iym, Buyukdag:2018ose, Buyukdag:2018cka, Smolkovic:2019jow, King:2020qaj, Arias-Aragon:2020bzy, Patel:2020bwo, Fedele:2020fvh} for recent attempts based on those ideas).

The clockwork mechanism~\cite{Giudice:2016yja} is another idea that generates exponentially small interactions in theories without small parameters at the fundamental level. As discussed in~\cite{Giudice:2016yja, Patel:2017pct, Ibarra:2017tju, Alonso:2018bcg, Hong:2019bki, deSouza:2019wji, vonGersdorff:2020ods, Kang:2020cxo, Babu:2020tnf}, a flavor clockwork can be constructed to explain the hierarchical flavor structure of the Yukawa couplings of the SM quarks and leptons. The clockwork mechanism has also useful applications in the context of axion physics~\cite{Farina:2016tgd, Coy:2017yex, Agrawal:2017eqm, Long:2018nsl, Darme:2020gyx, Bonnefoy:2018ibr}, inflation~\cite{Kehagias:2016kzt, Tangarife:2017rgl,Im:2017eju}, and dark sector physics~\cite{Hambye:2016qkf, Lee:2017fin, Goudelis:2018xqi, Craig:2018yld, Gherghetta:2019coi, Bae:2020hys}.
Additional developments are discussed in~\cite{Ahmed:2016viu, Craig:2017cda, Kehagias:2017grx, Teresi:2018eai, Sannino:2019sch, Darme:2021cxx}.

The flavor clockwork mechanism introduces, for each quark and lepton of the SM, a system of fermionic ``gears'', consisting of $N$ chiral fermions and $N+1$ fermions with opposite chirality. By introducing a specific set of mass terms for the fermions, one finds a single massless mode for each fermion flavor, with the remaining fields combining into $N$ massive Dirac fermions. The massless mode contains a component of the last site of the clockwork that is increasingly exponentially suppressed as the number of gears, $N$, rises. Any interaction of fields with the last site thereby translates into exponentially suppressed couplings of the massless modes. 
In the flavor clockwork setup, the massless modes are identified with the fermions of the SM. The gear number $N$ is chosen for each flavor to arrive at appropriately suppressed Yukawa couplings to the Higgs boson. 
As long as the \CW mass scale is well above the SM scale of electroweak symmetry breaking, the massive clockwork gear fermions remain unobservable. For a sufficiently small \CW scale, they might be collider accessible~\cite{Alonso:2018bcg}. 

Despite the absence of any sign of supersymmetric particles at the LHC~\cite{Canepa:2019hph}, supersymmetry (SUSY) remains one of the best motivated extensions of the SM. In the context of a flavor clockwork setup, SUSY removes the sensitivity of the electroweak scale to the clockwork scale, stabilizing the Higgs mass against loop corrections from the massive clockwork gears. In a SUSY version of the flavor clockwork, the fermionic gears of the clockwork are promoted to chiral superfields and the clockwork is implemented at the level of the superpotential. Such a setup results in hierarchical Yukawa couplings for the zero modes that are identified with the fermions and sfermions of the minimal supersymmetric standard model (MSSM). Once soft SUSY breaking is included, the clockwork mechanism will not only generate a hierarchical spectrum of quarks and leptons, but will also leave characteristic imprints on the spectrum of the squarks and sleptons. In turn, the squarks and sleptons can give indirect signatures in low-energy flavor violating processes.

It is well known that Flavor Changing Neutral Current (FCNC) processes, that are forbidden in the SM at tree level, are highly sensitive probes of New Physics (NP). Particularly sensitive processes are meson oscillations that can probe generic flavor violating NP at scales as high as $\Lambda_\text{NP} \sim 10^{5}$\,TeV~\cite{Isidori:2010kg,Silvestrini:2018dos,Aebischer:2020dsw}.
In this paper, we show that in a SUSY flavor clockwork setup, sizable flavor mixing among first and second generation squarks is predicted if SUSY breaking is gravity mediated. We consider constraints from Kaon mixing and derive generic bounds on the masses of squarks and gluinos. We find that the Kaon mixing constraints force either squarks or gluinos to have masses of at least $\sim 3$ PeV, introducing a considerable hierarchy between the electroweak scale and the scale of SUSY breaking masses.

This paper is organized as follows: In section~\ref{Sec:CWSM} we provide the outline of a clockwork flavor model. The model is subsequently supersymmetrized in section~\ref{Sec:SUSYCW}, where we detail the predicted flavor structure of the sfermion masses. The constraints on the SUSY breaking masses from Kaon mixing are discussed in section~\ref{sec:bounds}. We conclude in section~\ref{sec:conclusions}.

\section{The Flavor Clockwork for \SM Fermions}\label{Sec:CWSM}

It is instructive to first review the setup of ``clockworking'' a single fermion, for the mechanism can easily be extended to any number of them. We follow~\cite{Giudice:2016yja}. Consider $N$ left-chiral fermions $\psi_L^{(A)}$ and $N+1$ fermions with opposite chirality, $\psi_R^{(A)}$. In the limit of vanishing fermion masses, a large global chiral symmetry emerges. Each fermion can be associated with an abelian $U(1)$ factor, labeled $U(1)_L^{(A)}$ and $U(1)_R^{(A)}$ for $\psi_L^{(A)}$ and $\psi_R^{(A)}$, respectively. Away from the massless limit, the global chiral symmetry is broken. The degree of symmetry breaking can be tracked by promoting the fermion masses to spurion fields charged under the $U(1)$ factors. 
The clockwork mechanism is realized if the only symmetry breaking spurions are masses $m^{(A)}$, with charges $(+1,-1)$ under $U(1)_L^{(A)} \times U(1)_R^{(A)}$ and masses $m_\chi^{(A)}$, with charges $(+1,-1)$ under $U(1)_L^{(A)} \times U(1)_R^{(A+1)}$. Taking for simplicity $m^{(A)} = m$ and $m_\chi ^{(A)} = - m \chi$, with a common clockwork mass scale $m$ and a gear ratio $|\chi| > 1$, the relevant part of the fermionic ``gear'' Lagrangian reads~\cite{Giudice:2016yja}

\begin{equation} \label{eq:CWSimpleLagr}
    \lagr_m = - m \sum_{A=1}^{N}  \left( \overline{\psi}_L^{(A)} \psi_R^{(A)} - \chi   \overline{\psi}_L^{(A)} \psi_R^{(A+1)} + \mathrm{h.c.} \right)\,.
\end{equation}
This corresponds to a $N \times (N+1)$ mass matrix in clockwork space
\begin{equation}
    \bm{m} = m \begin{pmatrix} 1 & -\chi & 0 & 0 & \cdots & 0 & 0 \\
    0 & 1 & -\chi & 0 & \cdots & 0 & 0 \\
    0 & 0 & 1 & -\chi & \cdots & 0 & 0 \\
    \vdots & \vdots & \vdots & \ddots & \ddots & \vdots & \vdots \\
    0 & 0 & 0 & \cdots & 1 & -\chi & 0\\
    0 & 0 & 0 & \cdots & 0 & 1 & -\chi
    \end{pmatrix} .
\end{equation}
The above mass matrix leaves a single chiral $U(1)_R$ symmetry unbroken. The residual chiral symmetry is characterized by the existence of a massless mode. This is made more apparent upon observing that the mass matrix annihilates the mode
\begin{align} \label{eq:MasslessMode}
    \xi_R &\equiv \mathcal{N}\sum_{A = 1}^{N+1} \frac{1}{\chi^{A-1}}\psi_R^{(A)} ~, \\
    \nonumber
    \mathcal{N}^2 &=  \frac{|\chi|^2 - 1}{|\chi|^2 - |\chi|^{-2 N}} \simeq 1 + \mathcal{O}\left(|\chi|^{-2}\right),
\end{align}
where $\mathcal{N}$ is a normalization
constant. As an eigenstate of the only unbroken part of the chiral symmetry, this is the only massless mode, as a singular value decomposition of the mass matrix confirms. All other mass eigenstates have non-degenerate masses of the order of the clockwork scale $m$.
Due to the unitarity of the diagonalization matrices, the projection of $\psi_R^{(N+1)}$ onto the massless mode $\xi_R$ is $(\mathcal{N} \chi^{-N})^*$.

We now review an example in which the clockwork mechanism leads to exponentially suppressed couplings of the massless mode to another sector. Consider an interaction of the last fermion gear $\psi_R^{(N+1)}$ modeled after the \SM Yukawa terms
\begin{equation}
    \lagr \supset - \tilde{Y} \overline{q}_L H \psi_R^{(N+1)} ~,
\end{equation}
where $q_L$ is a left-handed SM quark, and $H$ the SM Higgs doublet. Note that this implies that all the clockwork fermions have to have the quantum numbers of the \SM right-handed up quark, $u_R = (3, 1, \frac{2}{3})$. The coupling $\Tilde{Y}$ is an $\ord{1}$ complex number, representing a proto-Yukawa coupling. Given the projection of $\psi_R^{(N+1)}$ into the massless mode $\xi_R$, it is apparent that the resulting Yukawa coupling $Y$ of the massless mode is exponentially suppressed by a factor $\chi^{-N}$
\begin{equation}\label{eq:ClockworkSupperssion}
    \lagr \supset - Y \overline{q}_L H \xi_R = - \left(\frac{\mathcal{N}}{\chi^N}\right)^*\tilde{Y} \overline{q}_L H \xi_R ~.
\end{equation}
In the following, we set the mass of all the massive modes in the clockwork (that have acted as ``gears'') far higher than the Higgs vacuum expectation value (vev), $v = 246$\,GeV, by choosing the clockwork scale $m \gg v$. In fact, we will assume that the massive fermions are sufficiently heavy, such that they have a negligible impact on phenomenology. The only observable state is the massless mode which is identified with the \SM right-handed up quark, $u_R$, and which has strongly suppressed interactions as specified in~\eqref{eq:ClockworkSupperssion}.
A similar construction can be applied to all other SM fermions as well.

Based on the discussion above, it is clear that the flavor structure of the SM Yukawa couplings may be attributed to a clockwork theory that generically gives exponential hierarchies in the couplings of the massless modes of the clockwork. 
Extending the mechanism to the entire Standard Model fermion content (see e.g.~\cite{Alonso:2018bcg}) corresponds to systematically assigning each fermion to the massless mode of its own gear system, with a characteristic clockwork mass scale that we assume to be far above the electroweak scale. In the following, we focus on the quark sector. A completely analogous construction can also be introduced for the charged leptons.

A crucial parameter that determines the couplings of the massless modes is the gear ratio $\chi$, which is generically expected to be an $\ord{1}$ number. In the context of quark Yukawa hierarchies, a natural choice is to relate $\chi$ to the sine of the Cabibbo angle, $\lambda$, such that $\chi = 1/\lambda \simeq 4$.
We consider the simplest possible clockwork extension of the \SM flavor sector, with a universal clockwork mass $m$ for all quark gear systems and a universal gear ratio $\chi$ 
\begin{align}
    \label{eq:ClockworkConstraints}
    m^{(A)}_{q^j_L} = m^{(A)}_{u^j_R} &= m^{(A)}_{d^j_R} = m ~,\\
    m\chi^{(A)}_{q^j_L} = m\chi^{(A)}_{u^j_R} = m &\chi^{(A)}_{d^j_R} = - m\chi = - \frac{m}{\lambda} ~.
\end{align}
The number of massive gears will be denoted by $\{N_q^j, N_u^j, N_d^j \}$, where $j$ runs over the flavor indices, $j = 1,2,3$. $N_q^j$ is the number of gears of the left-handed quark doublets of flavor $j$, while $N_u^j$ and $N_d^j$ are the numbers of gears of the right-handed up quark and down quark singlets of flavor $j$. 

These parameters, as well as the choice of proto-Yukawa couplings determine the Yukawa couplings of the SM quarks. Following the same procedure as in the single fermion case discussed above, one finds the SM Yukawa couplings
\begin{equation}\label{eq:YukawaCW}
    Y_x^{ij} \simeq \tilde{Y}_x^{ij} ~\chi^{-N_q^i-N_x^j}
    ~, \quad x = u,d ~.
\end{equation}
Here and in the following we work to leading order in an expansion in the gear ratio $1/\chi = \lambda$.
Diagonalizing the Yukawa matrices in~\eqref{eq:YukawaCW} determines the quark masses and the Cabibbo-Kobayashi-Maskawa (CKM) quark mixing matrix
\begin{align}
    \text{diag}(m_u,m_c,m_t) &= \frac{v}{\sqrt{2}} V_{L_u} Y_u V_{R_u}^\dagger~, \\
    \text{diag}(m_d,m_s,m_b) &= \frac{v}{\sqrt{2}} V_{L_d} Y_d V_{R_d}^\dagger~, \\
    V_\text{CKM} &= V_{L_u} V_{L_d}^\dagger ~.
\end{align}
For the diagonalization matrices one finds the following scaling with the gear ratio
\begin{align}\label{eq:ApproximateYukawaRotation}
    V_{L_x}^{ij} \sim \chi^{-|N_q^j-N_q^i|} ,~~ V_{R_x}^{ij} \sim \chi^{-|N_x^j-N_x^i|} ,~~ x = u,d~.
\end{align}
Due to the suppression of the off-diagonal entries of the diagonalization matrices by powers of the gear ratio, the quark masses are, to a good approximation, given by the diagonal entries in the Yukawa couplings
\begin{align}
 m_u \simeq \frac{v}{\sqrt{2}} |\tilde Y_u^{11}| \lambda^{N_q^1 + N_u^1} ~,&~~ m_d \simeq \frac{v}{\sqrt{2}} |\tilde Y_d^{11}| \lambda^{N_q^1 + N_d^1} ~, \\
 m_c \simeq \frac{v}{\sqrt{2}} |\tilde Y_u^{22}| \lambda^{N_q^2 + N_u^2} ~,&~~ m_s \simeq \frac{v}{\sqrt{2}} |\tilde Y_d^{22}| \lambda^{N_q^2 + N_d^2} ~, \\
 m_t \simeq \frac{v}{\sqrt{2}} |\tilde Y_u^{33}| \lambda^{N_q^3 + N_u^3} ~,&~~ m_b \simeq \frac{v}{\sqrt{2}} |\tilde Y_d^{33}| \lambda^{N_q^3 + N_d^3} ~.
\end{align}
In the equations above, the quark masses should be evaluated at the clockwork scale. The renormalization group running to high scales can have a significant impact on the quark masses. Therefore, it is convenient to work with ratios of quark masses that are, to a reasonable approximation, renormalization group invariant~\cite{Altmannshofer:2014qha}. Using the quark masses from~\cite{Zyla:2020zbs} and evolving them to a common renormalization scale $\mu = v = 246$\,GeV, taking into account 3 loop QCD running~\cite{Chetyrkin:2000abc}, we find
\begin{eqnarray}
\frac{m_u}{m_t} \simeq 7.3 \times 10^{-6} ~,&&~~ \frac{m_d}{m_b} \simeq 9.5 \times 10^{-4} ~, \\
\frac{m_c}{m_t} \simeq 3.7 \times 10^{-3} ~,&&~~ \frac{m_s}{m_b} \simeq 1.9 \times 10^{-2}~,
\end{eqnarray}
\begin{equation}
\frac{m_b}{m_t} \simeq 1.7 \times 10^{-2} ~.
\end{equation}
Since the top quark mass is of the order of the electroweak scale and gear numbers are non-negative, we set
\begin{equation}\label{eq:topConstraints}
 N_q^3 = N_u^3 = 0~.
\end{equation}
Using $\lambda = 0.2252$~\cite{Zyla:2020zbs}, we find that the quark mass ratios give the following constraints on the other gear numbers
\begin{align}
    \label{eq:UpDownConstraints}
    N_q^1 + N_u^1 = 8 ~,~~ N_q^1 + N_d^1 - N_d^3 = 4 ~,\\
    \label{eq:CharmStrangeConstraints}
    N_q^2 + N_u^2 = 4 ~,~~ N_q^2 + N_d^2 - N_d^3 = 2 ~,\\
    \label{eq:TopBottomConstraints}
    N_d^3 = 3 ~,
\end{align}
where we pick the integer gear numbers such as to minimize the deviation of the relevant proto-Yukawa ratios from $1$.

In the case of the CKM matrix, it is sufficient to consider the scaling with powers of $\lambda$
\begin{equation}
    V_\text{CKM} \sim \begin{pmatrix} 1 & \lambda^{N_q^1-N_q^2} & \lambda^{N_q^1-N_q^3} \\
    \lambda^{N_q^1-N_q^2} & 1 & \lambda^{N_q^2-N_q^3} \\
    \lambda^{N_q^1-N_q^3} & \lambda^{N_q^2-N_q^3} & 1 \end{pmatrix} .
\end{equation}
Comparing with the usual Wolfenstein parameterization, gives two additional conditions on the gear numbers of the left-handed quark doublets
\begin{equation}\label{eq:CKMConstraint}
    N_q^1 - N_q^3 = 3 ~,\quad
    N_q^2 - N_q^3 = 2~.
\end{equation}
The complex off-diagonal entries of the $\ord{1}$ proto-Yukawa couplings can be adjusted to reproduce the known magnitudes of the CKM matrix elements and the CP-violating CKM phase. Analogous to the quark masses, also the CKM matrix elements should be evaluated at the clockwork scale. However, the scale dependence of the CKM matrix is very mild~\cite{Balzereit:1998id} and can be neglected for our purposes.

The conditions~\eqref{eq:topConstraints}-\eqref{eq:TopBottomConstraints} combined with~\eqref{eq:CKMConstraint} have the solution
\begin{equation}\label{eq:FlavorCWGearNumbers}
    N_q = \left(
\begin{array}{c}
 3 \\
 2 \\
 0 \\
\end{array}
\right), \quad
N_u = \left(
\begin{array}{c}
 5 \\
 2 \\
 0 \\
\end{array}
\right), \quad
N_d = \left(
\begin{array}{c}
 4 \\
 3 \\
 3 \\
\end{array}
\right)~.
\end{equation}
These gear numbers correspond to the following scaling of the Yukawa couplings with the Cabibbo angle
\begin{equation}
    Y_u \sim \begin{pmatrix}
    \lambda^8 & \lambda^5 & \lambda^3 \\
    \lambda^7 & \lambda^4 & \lambda^2 \\
    \lambda^5 & \lambda^2 & 1 \end{pmatrix} ~,~~~ Y_d \sim \begin{pmatrix}
    \lambda^7 & \lambda^6 & \lambda^6 \\
    \lambda^6 & \lambda^5 & \lambda^5 \\
    \lambda^4 & \lambda^3 & \lambda^3 \end{pmatrix} ~,
\end{equation}
where we do not explicitly show the dependence on the proto-Yukawa couplings.
For a similar analysis, one may also follow the discussion in~\cite{Alonso:2018bcg} where a slightly different choice of gear numbers has been made (see also~\cite{Leurer:1993gy}). Small changes in gear numbers can be compensated by corresponding changes in the proto-Yukawa couplings and the precise value of the gear ratio $\chi$.

The above discussion serves to highlight the validity and key features of the \CW mechanism in the context of a setup that explains the \SM flavor hierarchies. 
In the next section, we will construct a supersymmetric version of this setup.


\section{The Supersymmetric Flavor Clockwork}\label{Sec:SUSYCW}


Starting from the flavor clockwork setup discussed above, we promote the clockwork fermions to chiral superfields. The matter field content then consists of gear systems for the three generations of left-handed quark doublet superfields $\bm{\Phi}_{Q_i}$, $\bm{\Phi}_{Q^c_i}$ the three generations of right-handed up-quark singlet superfields $\bm{\Phi}_{U^c_i}$, $\bm{\Phi}_{U_i}$, and the three generations of right-handed down-quark singlet superfields $\bm{\Phi}_{D^c_i}$, $\bm{\Phi}_{D_i}$, where $i = 1,2,3$ is the flavor index. The boldface notation indicates that the superfields are vectors in clockwork space, each with an individual multiplicity of fields. The field content with the respective multiplicities is summarized in Table~\ref{tab:SuperfieldContent}.

\setlength{\tabcolsep}{10pt}
\renewcommand{\arraystretch}{1.7}
\begin{table}
\begin{tabular}{ Oc Oc Oc }\hline\hline
    Superfield   & Gauge Representations                     & Multiplicity \\ 
    \hline
    $\bm{\Phi}_{Q_i}$   & $(3,2,\frac{1}{6})$                         & $N_q^i + 1$         \\
    $\bm{\Phi}_{Q^c_i}$ & $(\overline{3},\overline{2}, -\frac{1}{6})$ & $N_q^i$              \\ \hline
    $\bm{\Phi}_{U_i}$   & $(3,1,\frac{2}{3})$                         & $N_u^i$         \\ 
    $\bm{\Phi}_{U^c_i}$ & $(\overline{3},1, -\frac{2}{3})$ & $N_u^i+1$             \\ \hline
    $\bm{\Phi}_{D_i}$   & $(3,1,-\frac{1}{3})$                         & $N_d^i$         \\ 
    $\bm{\Phi}_{D^c_i}$ & $(\overline{3},1, \frac{1}{3})$ & $N_d^i+1$\\ \hline\hline             
\end{tabular}
\caption{Field content of the supersymmetric flavor clockwork. The multiplicities indicate the vector dimension of the corresponding superfield in clockwork space.}
\label{tab:SuperfieldContent}
\end{table}
\renewcommand{\arraystretch}{1.2}

Including the usual two Higgs doublets of the MSSM, $H_u$ and $H_d$, the relevant part of the superpotential reads
\begin{multline}\label{eq:Superpotential}
    W \supset \bm{\Phi}_{Q_i} \bm{m}_{Q_i} \bm{\Phi}_{Q_i^c} + \bm{\Phi}_{U_i} \bm{m}_{U_i} \bm{\Phi}_{U_i^c} + \bm{\Phi}_{D_i} \bm{m}_{D_i} \bm{\Phi}_{D_i^c}  \\
    + \tilde{Y}_u^{ij} H_u \bm{\Phi}_{Q_i} \bm{\Delta}_u^{ij} \bm{\Phi}_{U_j^c} + \tilde{Y}_d^{ij} H_d \bm{\Phi}_{Q_i} \bm{\Delta}_d^{ij} \bm{\Phi}_{D_j^c} ~,
\end{multline}
where sums over the flavor indices $i,j$ remain implicit. In addition, the superpotential also includes the standard Peccei-Quinn breaking term of the Higgs doublets, $\mu H_u H_d$. The clockwork masses $\bm{m}_{Q_i}$, $\bm{m}_{U_i}$, and $\bm{m}_{D_i}$ are $(N_q^i+1)\times N_q^i$, $N_u^i\times (N_u^i+1)$, and $N_d^i\times (N_d^i+1)$ matrices in clockwork space, and they are multiplied by superfield vectors of corresponding size. Assuming for simplicity that the mass matrices of the clockwork structure are universal, with a common clockwork scale $m = \Lambda_\text{CW}$ and a common gear ratio $\chi = 1/\lambda$ one has 
\begin{align}
    \label{eq:SuperpotentialxMassMatrix}
    (\bm{m}_{X_i})_{AB} &= m \left( \delta^{A}_{B} - \chi \delta^{A+1}_{B} \right), \quad X=U,D ~, \\
    \label{eq:SuperpotentialqMassMatrix}
    (\bm{m}_{Q_i})_{AB} &= m \left( \delta^{A}_{B} - \chi \delta^{A}_{B+1} \right).
\end{align}
The $\bm{\Delta}_u^{ij}$ and $\bm{\Delta}_d^{ij}$ in the second line of~\eqref{eq:Superpotential} are $(N_q^i+1)\times(N_x^j+1)$ matrices in clockwork space with a single non-zero entry for the last sites of each gear system
\begin{equation}
(\bm{\Delta}_x^{ij})_{AB}  = \delta^{N_q^i+1}_A \delta^{N_x^j+1}_B ~, \quad x = u,d ~.
\end{equation}
The corresponding $\tilde Y_u$ and $\tilde Y_d$ trilinear coefficients are flavor anarchic proto-Yukawa couplings that connect the three flavors of up-quark and down-quark fields at the last site of each gear system.

Next, we introduce soft SUSY breaking terms to provide all superpartners of Standard Model fields with masses above the current bounds from direct searches. We assume that SUSY breaking is gravity mediated to the visible sector by Planck scale suppressed operators. As long as the clockwork scale is sufficiently below the Planck scale $\Lambda_\text{CW} < M_\text{Pl}$, the SUSY breaking Lagrangian respects the chiral symmetries of the clockwork structure, and mixing among different flavors is only possible at the last site of each gear system. Therefore, the relevant soft masses for the squarks can be written as
\begin{multline}\label{eq:SoftSUSYBreaking}
    \mathcal L_\text{soft} \supset \mathbf{\tilde Q}_i^\dagger \bm{\mu}^2_{Q_i} \mathbf{\tilde Q}_i + \mathbf{\tilde Q}_i^{c\,\dagger} \bm{\mu}^2_{Q^c_i} \mathbf{\tilde Q}_i^c + (M^2_Q)_{ij}\mathbf{\tilde Q}_i^\dagger \bm{\Delta}^{ij}_{Q} \mathbf{\tilde Q}_j \\
    + \mathbf{\tilde U}_i^\dagger \bm{\mu}^2_{U_i} \mathbf{\tilde U}_i + \mathbf{\tilde U}_i^{c\,\dagger} \bm{\mu}^2_{U^c_i} \mathbf{\tilde U}_i^c + (M^2_U)_{ij}\mathbf{\tilde U}_i^{c\,\dagger} \bm{\Delta}^{ij}_{U} \mathbf{\tilde U}_j^c \\
    + \mathbf{\tilde D}_i^\dagger \bm{\mu}^2_{D_i} \mathbf{\tilde D}_i + \mathbf{\tilde D}_i^{c\,\dagger} \bm{\mu}^2_{D^c_i} \mathbf{\tilde D}_i^c + (M^2_D)_{ij}\mathbf{\tilde D}_i^{c\,\dagger} \bm{\Delta}^{ij}_{D} \mathbf{\tilde D}_j^c ~,
\end{multline}
where sums over flavor indices $i,j$ are implicit, and  $\mathbf{\tilde Q}_i$, $\mathbf{\tilde Q}_i^c$, $\mathbf{\tilde U}_i$, $\mathbf{\tilde U}_i^c$, and $ \mathbf{\tilde D}_i$, $\mathbf{\tilde D}_i^c$ are the scalar components of the superfield gear systems. The various soft breaking masses $\bm{\mu}^2$ are diagonal matrices in clockwork space. Generically, we expect the non-zero entries to be of the same order that we denote with $m_S^2$.
Similar as in~\eqref{eq:Superpotential}, the $\mathbf{\Delta}_Q^{ij}$, $\mathbf{\Delta}_U^{ij}$, and $\mathbf{\Delta}_D^{ij}$ are $(N_q^i+1)\times(N_q^j+1)$, $(N_u^i+1)\times(N_u^j+1)$, and $(N_d^i+1)\times(N_d^j+1)$ matrices in clockwork space, respectively, each with a single non-zero entry for the last clockwork site. The soft breaking masses $(M^2_Q)_{ij}$, $(M^2_U)_{ij}$, and $(M^2_D)_{ij}$ are generically expected to be flavor anarchic and also of the order $\sim m_S^2$.
 
The trilinear $A$-terms that couple squarks to the Higgs bosons have a structure similar to the proto-Yukawa couplings in the superpotential
\begin{equation} \label{eq:trilinear}
    \mathcal L_\text{soft} \supset A_{u}^{ij} H_u \mathbf{\tilde Q}_{i} \bm{\Delta}_u^{ij} \mathbf{\tilde U}_{j}^c + A_{d}^{ij} H_d \mathbf{\tilde Q}_{i} \bm{\Delta}_d^{ij} \mathbf{\tilde D}_{j}^c ~+ \text{h.c.}~.
\end{equation}
Generically, one expects the trilinear terms to be flavor anarchic, and of the same order of the soft masses $A_{x}^{ij} \sim m_S$, or 1-loop suppressed $A_{x}^{ij} \sim m_S/16\pi^2$ if their dominant contribution arises from anomaly mediation~\cite{Giudice:1998xp}. The lack of experimental evidence for SUSY particles implies that the SUSY masses have to be significantly above the electroweak scale. In such a case, the trilinear terms have a negligible impact on the low energy flavor observables that we will analyze in the next section. Therefore, we will not consider the trilinear terms in any further detail.

In addition to the soft SUSY breaking terms discussed above, the gaugino masses and the $B\mu$ term can be introduced analogously to the MSSM. Of particular relevance for the discussion of the flavor constraints in section~\ref{sec:bounds} is the gluino mass $m_{\tilde g}$. We will assume that the gluino mass is of the order of the soft scalar masses $m_S$, keeping in mind that 1-loop suppressed gaugino masses are motivated as well in the context of anomaly mediation~\cite{Giudice:1998xp}.

The scenario we consider is then characterized by the following important scales
\begin{equation}
    v < m_{\tilde g}, m_S \ll \Lambda_\text{CW} < M_\text{Pl} ~. 
\end{equation}
As discussed in appendix~\ref{appendix}, the large number of matter fields in the clockwork setup significantly modifies the running of the gauge couplings above the clockwork scale and leads to a loss of asymptotic freedom. Requiring the absence of Landau poles below the Planck scale, $M_\text{Pl}$, implies a lower bound on the clockwork scale a few orders of magnitude below $M_\text{Pl}$.
SUSY stabilizes the electroweak scale against quantum corrections of the order of the clockwork scale or the Planck scale.  As SUSY holds to an excellent approximation at the clockwork scale, all superfields $\bm{\Phi}$ may be rotated into their mass eigenbases that render the clockwork masses in~\eqref{eq:SuperpotentialxMassMatrix} and~\eqref{eq:SuperpotentialqMassMatrix} diagonal. This leads to $N_q^i$, $N_u^i$, and $N_d^i$ massive modes with masses of the order of the clockwork scale $\Lambda_\text{CW}$, that are decoupled from phenomenology. The massless modes may consequently be identified with the quarks and squarks of the MSSM $(q_L^i,\tilde{q}_L^i)$, $(u_R^i,\tilde{u}^i_R)$, and $(d_R^i,\tilde{d}^i_R)$. Analogous to the non-SUSY case given in~\eqref{eq:MasslessMode} we find
\begin{align}
    \label{eq:LHSquarkModes}
    (q_L^i,\tilde{q}_L^i) &\simeq
    \sum_{A = 1}^{N_q^i+1} \frac{1}{\chi^{A-1}} ( \mathbf{Q}_i, \mathbf{\tilde Q}_i)_A ~,\\ 
    \label{eq:RHSquarkModes}
    (u_R^{i\,c},\tilde{u}_R^{i\,*}) & \simeq 
    \sum_{A = 1}^{N_u^i+1} \frac{1}{\chi^{A-1}}( \mathbf{U}^c_i, \mathbf{\tilde U}^c_i)_A ~,\\
    (d_R^{i\,c},\tilde{d}_R^{i\,*}) & \simeq 
    \sum_{A = 1}^{N_d^i+1} \frac{1}{\chi^{A-1}}( \mathbf{D}^c_i, \mathbf{\tilde D}^c_i)_A ~,
\end{align}
where we suppressed normalization factors of order  $1 + \ord{\chi^{-2}}$.
The Yukawa couplings of the massless superfields have the exponential structure given in~\eqref{eq:YukawaCW}.

In a supersymmetric context, the gear numbers derived in \eqref{eq:FlavorCWGearNumbers} need to be adjusted. As up-type quarks obtain their mass from the Higgs doublet $H_u$ and down-type quarks from the Higgs doublet $H_d$, part of the hierarchical fermion spectrum could be due to a large ratio of Higgs vevs $\tan\beta = v_u/v_d$. Among the constraints on the gear numbers discussed in section~\ref{Sec:CWSM}, the constraint in~\eqref{eq:TopBottomConstraints} gets therefore modified to
\begin{equation}
   N_d^3 = 3 - n ~, 
\end{equation}
where the integer $n$ is determined by the size of $\tan\beta$ in terms of the gear ratio $\tan\beta \sim \chi^n$. Viable choices are $n = 0,1,2,3$. The resulting gear numbers are
\begin{equation}\label{eq:SUSYFlavorCWGearNumbers}
    N_q = \left(
\begin{array}{c}
 3 \\
 2 \\
 0 \\
\end{array}
\right), ~~
N_u = \left(
\begin{array}{c}
 5 \\
 2 \\
 0 \\
\end{array}
\right), ~~
N_d = \left(
\begin{array}{c}
 4-n \\
 3-n \\
 3-n \\
\end{array}
\right)~.
\end{equation}

The squark components of the massless superfields acquire soft SUSY breaking masses from the terms in~\eqref{eq:SoftSUSYBreaking}. Since the clockwork mass scale, $\Lambda_\text{CW}$, is much larger than the soft scalar masses $\sim m_S$, the treatment of the soft masses as a perturbation on the clockwork structures provides an excellent approximation of the squark mass matrices. We find for the left-handed and right-handed  squark soft masses of the MSSM
\begin{align} \label{eq:FlavorLHSquarkMasses}
    (\hat m^2_{\tilde q_L})_{ij} &\simeq (\bm{\mu}^2_{Q_i})_{11} \delta_{ij}  + (M^2_Q)_{ij} \chi^{-N_q^i-N_q^j} ~, \\
    \label{eq:FlavorRHuSquarkMasses}
    (\hat m^2_{\tilde u_R})_{ij} &\simeq  (\bm{\mu}^2_{U_i})_{11} \delta_{ij}  + (M^2_U)^*_{ji} \chi^{-N_u^i-N_u^j} ~,\\
    \label{eq:FlavorRHdSquarkMasses}
    (\hat m^2_{\tilde d_R})_{ij} &\simeq  (\bm{\mu}^2_{D_i})_{11} \delta_{ij}  + (M^2_D)^*_{ji} \chi^{-N_d^i-N_d^j} ~.
\end{align}
These mass matrices are obtained upon discarding terms that are higher order in $m_S^2/\Lambda^2_\text{CW}$ and in the gear ratio $1/\chi$. The leading flavor diagonal squark masses are provided by the soft SUSY breaking terms of the first clockwork gears $(\bm{\mu}^2_{Q_i})_{11}$, $(\bm{\mu}^2_{U_i})_{11}$, and $(\bm{\mu}^2_{D_i})_{11}$. These masses are all of order $m_S^2$ and not suppressed by the gear ratio. It is important to note that these masses are flavor diagonal but not necessarily flavor universal. The flavor violating masses originate from the $M_Q^2$, $M_U^2$, and $M_D^2$ terms and are suppressed by powers of the gear ratio. The generic scaling with the sine of the Cabibbo angle is given by
\begin{align} \label{eq:mqL}
    \hat m^2_{\tilde q_L} \sim m_S^2 \begin{pmatrix} 1 & \lambda^{N_q^1+N_q^2} & \lambda^{N_q^1+N_q^3} \\ \lambda^{N_q^1+N_q^2} & 1 & \lambda^{N_q^2+N_q^3} \\
    \lambda^{N_q^1+N_q^3} & \lambda^{N_q^2+N_q^3} & 1
    \end{pmatrix} , \\ \label{eq:mxR}
    \hat m^2_{\tilde x_R} \sim m_S^2 \begin{pmatrix} 1 & \lambda^{N_x^1+N_x^2} & \lambda^{N_x^1+N_x^3} \\ \lambda^{N_x^1+N_x^2} & 1 & \lambda^{N_x^2+N_x^3} \\
    \lambda^{N_x^1+N_x^3} & \lambda^{N_x^2+N_x^3} & 1
    \end{pmatrix} ,
\end{align}
where $x = u,d$.
At first sight, this flavor structure of the squark masses seems to indicate strongly suppressed flavor mixing, in particular between the first and second generation of squarks.
However, to make the connection to phenomenology, it is important to express the squark masses in~\eqref{eq:mqL} and~\eqref{eq:mxR} in the super-CKM basis. Performing the flavor rotations~\eqref{eq:ApproximateYukawaRotation} that diagonalize the quark Yukawa couplings also on the squarks gives the following squark mass matrices
\begin{align}
        m^2_{\tilde x_L} &= V_{L_x} \hat m^2_{\tilde q_L} V_{L_x}^\dagger , \quad x=u,d ~, \\
        m^2_{\tilde x_R} &= V_{R_x} \hat m^2_{\tilde x_R} V_{R_x}^\dagger , \quad x=u,d ~,
\end{align}
where the un-hatted $m^2_{\tilde x_R}$ and $m^2_{\tilde x_L}$ are the squark mass matrices in the super-CKM basis. As mentioned above, the diagonal entries of $\hat m^2_{\tilde x_R}$ and $\hat m^2_{\tilde q_L}$ are not universal and the rotations therefore tend to increase the off-diagonal entries. This is particularly relevant for the $1\leftrightarrow2$ mixing entries. We find
\begin{align}
    m^2_{\tilde x_L} \sim m_S^2 \begin{pmatrix} 1 & \lambda^{|N_q^1-N_q^2|} & \lambda^{|N_q^1-N_q^3|} \\ \lambda^{|N_q^1-N_q^2|} & 1 & \lambda^{|N_q^2-N_q^3|} \\
    \lambda^{|N_q^1-N_q^3|} & \lambda^{|N_q^2-N_q^3|} & 1
    \end{pmatrix} , \\
    m^2_{\tilde x_R} \sim m_S^2 \begin{pmatrix} 1 & \lambda^{|N_x^1-N_x^2|} & \lambda^{|N_x^1-N_x^3|} \\ \lambda^{|N_x^1-N_x^2|} & 1 & \lambda^{|N_x^2-N_x^3|} \\
    \lambda^{|N_x^1-N_x^3|} & \lambda^{|N_x^2-N_x^3|} & 1
    \end{pmatrix} ,
\end{align}
where $x = u,d$.
Note that the mass matrices of the right-handed up and down squarks, $m^2_{\tilde u_R}$ and $m^2_{\tilde d_R}$, are independent from each other. In contrast, the mass matrices for the left-handed squarks are related due to $SU(2)_L$ symmetry 
\begin{equation}
   m_{\tilde u_L}^2 = V_\text{CKM } m_{\tilde d_L}^2 V_\text{CKM}^\dagger ~. 
\end{equation}
Using the gear numbers in~\eqref{eq:SUSYFlavorCWGearNumbers}, the squark masses in the super-CKM basis have the following explicit scaling with the Cabibbo angle, independently of the value for $\tan\beta$
\begin{equation} \label{eq:LLsquark_mass}
    \frac{m_{\tilde d_L}^2}{m_S^2} \sim  \begin{pmatrix}
    1 & \lambda & \lambda^3 \\
    \lambda & 1 & \lambda^2 \\
    \lambda^3 & \lambda^2 & 1 \end{pmatrix} ~,
\end{equation}
\begin{equation} \label{eq:RRsquark_mass}
    \frac{m_{\tilde u_R}^2}{m_S^2} \sim  \begin{pmatrix}
    1 & \lambda^3 & \lambda^5 \\
    \lambda^3 & 1 & \lambda^2 \\
    \lambda^5 & \lambda^2 & 1 \end{pmatrix} ~,~~~ \frac{m_{\tilde d_R}^2}{m_S^2} \sim \begin{pmatrix}
    1 & \lambda & \lambda \\
    \lambda & 1 & 1 \\
    \lambda & 1 & 1 \end{pmatrix} ~.
\end{equation}
The flavor diagonal entries are all of the same order, but not universal. The off-diagonal entries can be complex and are generically expected to have $\mathcal O(1)$ CP-violating phases. We note in particular sizeable entries of $\mathcal O (\lambda)$ that lead to mixing of the first and second generation of down type squarks. This is reminiscent of the situation in the simplest supersymmetric Froggatt-Nielsen flavor models with a single $U(1)$ flavor symmetry~\cite{Leurer:1993gy}.

It could be interesting to explore variations of our setup that lead to hierarchical sfermion masses as in~\cite{Nomura:2008pt}, or to a sfermion spectrum with inverted hierarchy as in~\cite{Buyukdag:2018ose,Buyukdag:2018cka}.

\section{Constraints from Meson Mixing} \label{sec:bounds}

In the presence of flavor violating squark masses, many low-energy flavor observables receive SUSY contributions \cite{Gabbiani:1996hi, Altmannshofer:2009ne}. Particularly sensitive probes of flavor violating squarks are Kaon oscillations that are potentially sensitive to squarks at the PeV scale~\cite{Altmannshofer:2013lfa, Isidori:2019pae}.

The observables of interest are $\Delta m_K$, the frequency of $K^0 - \bar{K}^0$ oscillations, and $\epsilon_K$, that measures indirect CP-violation $K^0 - \bar{K}^0$ oscillations. The computation of meson oscillation observables is commonly cast into an OPE framework, providing an effective Hamiltonian consisting of dimension 6 operators $O_k$ and their corresponding Wilson coefficients $C_k$
\begin{equation}
    \mathcal H_\text{eff} = \sum_{k=1}^5 C_k O_k + \sum_{k=1}^3 \tilde C_k \tilde O_k  + \text{h.c.}~.  
\end{equation}
Based on this effective Hamiltonian, the observables can be computed as
\begin{align}\label{eq:DeltamK}
    \Delta m_K &=  2\, \mathrm{Re} \bra{K^0} \mathcal{H}_{\mathrm{eff}} \ket{\bar{K}^0},\\ \label{eq:epsilonK}
    |\epsilon_K| &= \frac{\kappa_\epsilon}{\sqrt{2}\Delta m_K} \mathrm{Im} \bra{K^0} \mathcal{H}_{\mathrm{eff}} \ket{\bar{K}^0},
\end{align}
where $\kappa_\epsilon \simeq 0.94 \pm 0.02$~\cite{Buras:2010pza} encapsulates long-distance contributions to the CP violating parameter.

The dominant SUSY contributions to the Wilson coefficients are coming from gluino-squark loops and are proportional to the mixing between the first and second generation of down-type quarks. Additional chargino-squark and neutralino-squark loop  contributions have been considered for example in~\cite{Altmannshofer:2007cs,Crivellin:2010ys}. They can be relevant in regions of parameter space where gluino contributions are accidentally small. This is not the case in our scenario.

The relevant entries in the squark mass matrices~\eqref{eq:RRsquark_mass} and~\eqref{eq:LLsquark_mass} that enter the gluino contributions are suppressed by the sine of the Cabibbo angle. It is thus justified to work in the mass insertion approximation, expanding the SUSY contributions to first order in the left-left and right-right mass insertions
\begin{equation}
    (\delta_{12}^d)_{LL} = \frac{(m_{\tilde d_L}^2)_{12}}{m_S^2} \sim \lambda , 
    ~~(\delta_{12}^d)_{RR} = \frac{(m_{\tilde d_R}^2)_{12}}{m_S^2} \sim \lambda .
\end{equation}
Additional left-right mixing mass insertions are proportional to the trilinear soft A-terms introduced in~\eqref{eq:trilinear}. They are suppressed by the electroweak scale $(\delta^d_{ij})_{LR} \propto v/m_S$ and can be safely neglected. In that case, the only relevant operators in the effective Hamiltonian are
\begin{align}\label{eq:WilsonOp1}
    O_1 &= (\overline{d}_L^\alpha \gamma_\mu s_L^\alpha) (\overline{d}_L^\beta \gamma^\mu s_L^\beta)~,\\ \label{eq:WilsonOp1tilde}
    \tilde O_1 &= (\overline{d}_R^\alpha \gamma_\mu s_R^\alpha) (\overline{d}_R^\beta \gamma^\mu s_R^\beta)~,\\
    \label{eq:WilsonOp4}
    O_4 &= (\overline{d}_R^\alpha s_L^\alpha) (\overline{d}_L^\beta s_R^\beta)~,\\
    \label{eq:WilsonOp5}
    O_5 &= (\overline{d}_R^\alpha s_L^\beta) (\overline{d}_L^\beta s_R^{\alpha})~,
\end{align}
where $\alpha,\beta = 1,\dots,3$ are color indices. In the mass insertion approximation, the Wilson coefficients depend on the gluino mass $m_{\tilde g}$ and the common squark mass $m_S$. They read~\cite{Gabbiani:1996hi,Ciuchini:1998ix,Altmannshofer:2009ne}
\begin{align}
    C_1 & = \frac{1}{216} \frac{\alpha_s^2}{m_S^2} (\delta_{12}^d)^2_{LL} f_1(m_{\tilde g}^2/m_S^2) ~, \\
    \tilde C_1 & = \frac{1}{216} \frac{\alpha_s^2}{m_S^2} (\delta_{12}^d)^2_{RR} f_1(m_{\tilde g}^2/m_S^2) ~, \\ \label{eq:C4}
    C_4 & = -\frac{23}{180} \frac{\alpha_s^2}{m_S^2} (\delta_{12}^d)_{LL} (\delta_{12}^d)_{RR} f_4(m_{\tilde g}^2/m_S^2) ~, \\
    C_5 & = \frac{7}{540} \frac{\alpha_s^2}{m_S^2} (\delta_{12}^d)_{LL} (\delta_{12}^d)_{RR}  f_5(m_{\tilde g}^2/m_S^2) ~.
\end{align}
The loop functions are normalized such that $f_i(1) = 1$, and the explicit expressions are
\begin{multline}
    f_1(x) = \frac{2(11+144x+27x^2-2x^3)}{(1-x)^4} \\ + \frac{12x(13+17x)}{(1-x)^5} \log x ~,
\end{multline}
\begin{multline}
    f_4(x) = \frac{10(2-99x-54x^2+7x^3)}{23(1-x)^4} \\ - \frac{60x(5+19x)}{23(1-x)^5} \log x ~,
\end{multline}
\begin{multline}
    f_5(x) = \frac{10(10+117x+18x^2-x^3)}{7(1-x)^4} \\ + \frac{60x(11+13x)}{7(1-x)^5} \log x ~.
\end{multline}
The operator matrix elements $\langle O_i \rangle = \bra{K^0} O_i \ket{\bar{K}^0}$, relevant to evaluate the observables $\Delta m_K$ and $\epsilon_K$ in~\eqref{eq:DeltamK} and~\eqref{eq:epsilonK}, can be written as
\begin{align}\label{eq:TransitionWilsonO1}
    \langle O_1 \rangle = \langle \tilde O_1 \rangle &= \frac{1}{3} B_1 m_{K} f_K^2~,\\ \label{eq:TransitionWilsonO4}
    \langle O_4 \rangle &= \frac{1}{4} B_4 m_K f_K^2 \frac{m_K^2}{(m_d + m_s)^2}~,\\ \label{eq:TransitionWilsonO5}
    \langle O_5 \rangle &= \frac{1}{12} B_5 m_K f_K^2 \frac{m_K^2}{(m_d + m_s)^2} ~,
\end{align}
where $f_K = 155.7$\,MeV~\cite{Aoki:2019cca} is the Kaon decay constant.
For the so-called bag parameters $B_i$ we use the lattice determinations from~\cite{Carrasco:2015pra} $B_1 = 0.506 \pm 0.017$, $B_4 = 0.78\pm 0.05$, and $B_5 = 0.49\pm0.04$. Note that these parameters are renormalization scheme and scale dependent and the given values correspond to the $\overline{ \text{MS}}$ scheme at a low hadronic scale of $\mu_\text{low} = 3$\,GeV. The quark masses in the matrix elements \eqref{eq:TransitionWilsonO4} and \eqref{eq:TransitionWilsonO5} have to be evaluated in the same scheme and at the same scale. We use $m_s(\mu_\text{low}) \simeq 86$\,MeV and $m_d(\mu_\text{low}) \simeq 4.3 $\,MeV~\cite{Zyla:2020zbs,Chetyrkin:2000abc}. Note the chiral enhancement of the matrix elements $\langle O_4 \rangle$ and $\langle O_5 \rangle$ by the factor $m_K^2/(m_d + m_s)^2 \simeq 30$.

In contrast to the hadronic matrix elements, the Wilson coefficients discussed above are defined at a high renormalization scale that corresponds to the masses of the SUSY particles $\mu_\text{high} \sim m_S, m_{\tilde g}$.
In particular, the strong coupling constant in the Wilson coefficients is evaluated at that scale. 
The Wilson coefficients are evolved to the low scale $\mu_\text{low}$ by solving their renormalization group equations~\cite{Ciuchini:1997bw, Buras:2000if, Buras:2001ra}. At the low scale they are combined with the hadronic matrix elements to determine the observables $\Delta m_K$ and $\epsilon_K$. As we consider the leading order SUSY contributions to the Wilson coefficients, 1-loop running is appropriate. The low scale Wilson coefficients are thus given by
\begin{align}
C_1(\mu_\text{low}) &= \eta_4^{6/25} \eta_5^{6/23} \eta_6^{2/7} C_1(\mu_\text{high}) ~, \\
\tilde C_1(\mu_\text{low}) &= \eta_4^{6/25} \eta_5^{6/23} \eta_6^{2/7} \tilde C_1(\mu_\text{high}) ~, \\
C_4(\mu_\text{low}) &= \eta_4^{-24/25} \eta_5^{-24/23} \eta_6^{-8/7} C_4(\mu_\text{high}) \nonumber \\ \label{eq:running4}
& ~~~~~~~+ \frac{1-\eta_4^{27/25} \eta_5^{27/23} \eta_6^{9/7}}{3 \eta_4^{24/25} \eta_5^{24/23} \eta_6^{8/7}} C_5(\mu_\text{high})~,\\
C_5(\mu_\text{low}) &= \eta_4^{3/25} \eta_5^{3/23} \eta_6^{1/7} C_5(\mu_\text{high})~,
\end{align}
where the $\eta_i$ factors are ratios of the strong coupling constant evaluated at different scales in a 4, 5, or 6 flavor scheme as appropriate
\begin{equation}
    \eta_6 = \frac{\alpha_s^{(6)}(\mu_\text{high})}{\alpha_s^{(6)}(\mu_t)}, ~ \eta_5 = \frac{\alpha_s^{(5)}(\mu_t)}{\alpha_s^{(5)}(\mu_b)}, ~ \eta_4 = \frac{\alpha_s^{(4)}(\mu_b)}{\alpha_s^{(4)}(\mu_\text{low})}.
\end{equation}
We use the numerical values $\eta_4 \simeq 0.89$ and $\eta_5 \simeq 0.48$~\cite{Chetyrkin:2000abc}. We decide to evaluate $\eta_6$ at the larger of the squark mass and gluino mass $\mu_\text{high} = \text{max}(m_{\tilde g},m_S)$.

\begin{figure*}[tbh]
\begin{center}
\includegraphics[width=0.48\textwidth]{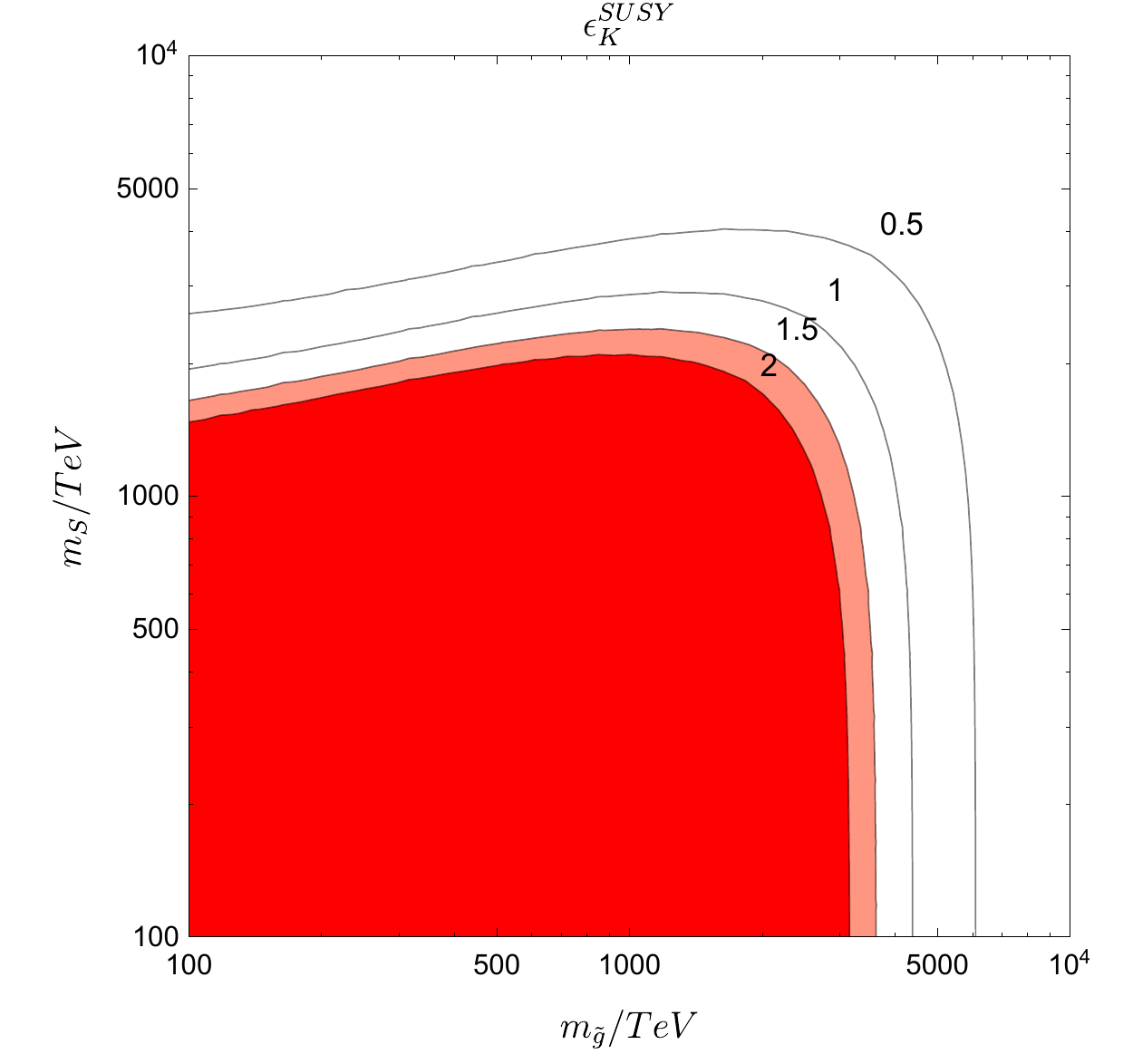}\enspace\enspace
\includegraphics[width=0.48\textwidth]{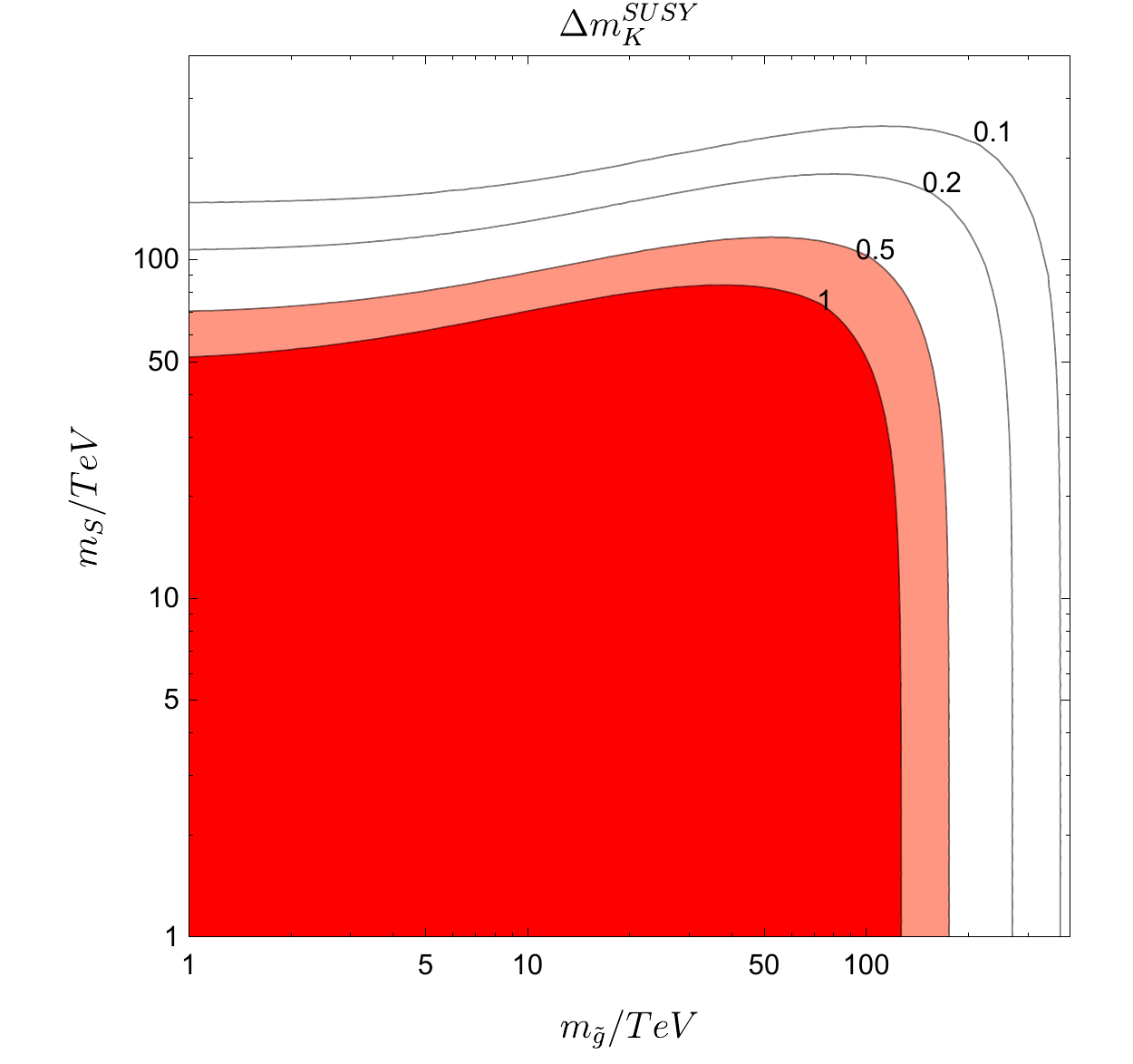}
\caption{Constraints on the squark and gluino masses from Kaon oscillations. In the red regions, the SUSY contributions to $\epsilon_K$ (left) and $\Delta m_K$ (right) exceed acceptable values. The solid black contours indicate the size of $\epsilon_K^\text{SUSY}$ relative to the SM uncertainty of $\epsilon_K$ and the size of $\Delta m_K^\text{SUSY}$ relative to the experimental central value of $\Delta m_K$, respectively.
}
\label{fig:ExclusionPlots}
\end{center}
\end{figure*}

In Figure~\ref{fig:ExclusionPlots}, we show the SUSY contributions to $\epsilon_K$ (left plot) and to $\Delta m_K$ (right plot) as function of the gluino mass $m_{\tilde g}$ and the generic squark mass $m_S$. 
The solid black contours in the left plot indicate the size of the contribution $\epsilon_K^\text{SUSY}$ relative to the uncertainty of the SM prediction~\cite{Brod:2019rzc}
\begin{equation}
    \epsilon_K^\text{SM} = (2.16 \pm 0.18) \times 10^{-3} ~,
\end{equation}
which is in excellent agreement with the very precise experimental value~\cite{Zyla:2020zbs}
\begin{equation}
    \epsilon_K^\text{exp} = (2.228 \pm 0.011) \times 10^{-3} ~.
\end{equation}
SUSY contributions significantly larger than the SM theory uncertainty are therefore excluded. The dark (light) red region in the left plot of Figure~\ref{fig:ExclusionPlots}, corresponds to a SUSY contribution larger than $2\times$ ($1.5\times$) the SM uncertainty. 

The solid contours in the right plot of Figure~\ref{fig:ExclusionPlots} indicate the size of $\Delta m_K^\text{SUSY}$ relative to the measured value~\cite{Zyla:2020zbs} 
\begin{equation}
    \Delta m_K^\text{exp} = (3.484 \pm 0.006) \times 10^{-15}~\text{GeV} ~.
\end{equation}
The SM prediction of $\Delta m_K$ is subject to large hadronic uncertainties. Perturbative calculations of the short-distance piece find values close to the experimental value with $\sim 40\%$ uncertainty~\cite{Brod:2011ty}. First lattice calculations that include both short-distance and long-distance parts have similar precision~\cite{Christ:2012se,Bai:2014cva,Wang:2020jpi}. In such a situation, it is a meaningful approach to bound the size of the SUSY contributions to $\Delta m_K$ by the experimental value itself. In the right plot of Figure~\ref{fig:ExclusionPlots}, the dark (light) red region corresponds to a $\Delta m_K^\text{SUSY}$ that saturates the experimental value (corresponds to $50\%$ of the experimental value). 

For definiteness we use the following values of the mass insertions in the plots
\begin{multline}
    \text{Re}\big[(\delta_{12}^d)_{LL}^2\big] = \text{Im}\big[(\delta_{12}^d)_{LL}^2\big] = \text{Re}\big[(\delta_{12}^d)_{RR}^2\big] = \\
    =\text{Im}\big[(\delta_{12}^d)_{RR}^2\big] =  -\text{Re}\big[(\delta_{12}^d)_{LL} (\delta_{12}^d)_{RR} \big] = \\ 
    = -\text{Im}\big[(\delta_{12}^d)_{LL} (\delta_{12}^d)_{RR} \big]= \lambda^2~.
\end{multline}
Order one changes in the mass insertions lead to order one changes in the excluded regions in the squark and gluino masses. Note that the new physics effect to Kaon mixing is dominated by the contribution from the Wilson coefficient $C_4\propto (\delta_{12}^d)_{LL} (\delta_{12}^d)_{RR}$, because $C_4$ contains the largest numerical pre-factor, see~\eqref{eq:C4}, and comes with a chirally enhanced matrix element $\langle O_4 \rangle$, see~\eqref{eq:TransitionWilsonO4}. The explicit choice of signs in the above mass insertions corresponds to a positive SUSY contribution to the Kaon mixing matrix element $M_{12}^K = \langle K^0 |\mathcal{H}_{\mathrm{eff}}| \bar K_0\rangle$. Different sign choices do not qualitatively change the results shown in Figure~\ref{fig:ExclusionPlots}. 

We see that $\epsilon_K$ severely constrains the masses of the SUSY particles. Either squarks or gluinos have to be heavier than a few PeV in the considered SUSY flavor clockwork model.
The constraint from $\Delta m_K$ is significantly weaker, leading to lower bounds on the squark or gluino masses of around 100 TeV.

\section{Discussion and Conclusions} \label{sec:conclusions}

In this paper, we constructed a supersymmetric version of a flavor model based on the clockwork mechanism. For each SM fermion, a gear system is introduced that leads to an exponential suppression of its Yukawa couplings with the exponent determined by the number of gear fermions. All massive gear fermions have a mass of the order of the clockwork scale $\Lambda_\text{CW}$ and decouple from phenomenology.  
SUSY stabilizes the electroweak scale against quantum corrections proportional to $\Lambda_\text{CW}$ (and the Planck scale). 

The flavor clockwork leaves a characteristic imprint on the masses of the fermions' superpartners. Assuming gravity mediated SUSY breaking, we find that the flavor clockwork predicts sizable flavor changing entries in the squark mass matrices. In particular, mixing between the first and second generation of squarks is predicted to be of the order of the Cabibbo angle. 

The large flavor mixing of squarks leads to large 1-loop contributions to Kaon oscillations. In the absence of accidental cancellations and assuming generic $\mathcal O(1)$ CP-violating phases, we find that squarks or gluinos have to have masses of at least a few PeV to avoid the stringent constraints from the Kaon mixing observable $\epsilon_K$.

For PeV scale SUSY particles, the new physics effects in other neutral meson systems is strongly suppressed. First, $B_s$, $B_d$, and $D^0$ meson oscillations are generically less sensitive to new physics compared to Kaon oscillations. Second, in the considered SUSY model, the relevant mass insertions in~\eqref{eq:LLsquark_mass} and~\eqref{eq:RRsquark_mass} are suppressed by additional powers of the Cabibbo angle $\lambda$. Taking into account only the dominant contribution from the Wilson coefficient $C_4$, and neglecting $\mathcal O(1)$ ratios of hadronic matrix elements, we find the following approximate scaling relations for the $B_s$ and $B_d$ mixing matrix elements
\begin{multline}
   \frac{(M_{12}^{B_s})_\text{SUSY}}{(M_{12}^{B_s})_\text{SM}}  \sim \frac{(M_{12}^{B_d})_\text{SUSY}}{(M_{12}^{B_d})_\text{SM}} \sim  \lambda^6 \frac{m_B^2}{m_K^2} \frac{m_s^2}{m_b^2} \frac{\epsilon_K^\text{SUSY}}{\epsilon_K^\text{SM}} \\ \lesssim  5\times 10^{-7}~,
\end{multline}
far below any foreseeable sensitivities. Similarly in $D^0$ mixing we expect new physics effects that are several orders of magnitude below the current sensitivities.

The mass of the SM-like Higgs boson of $m_h \simeq 125.5$\,GeV is easily achieved in a scenario with squarks at the PeV scale without the need of large stop mixing~\cite{Giudice:2011cg}. The preferred value for $\tan\beta$ is on the low side, $\tan\beta \sim \mathcal O(1)$ or $\mathcal O(\chi)$.

On the downside, very heavy squarks and gluinos correspond to a severe fine tuning of the electroweak scale of the order of $v^2 / m_S^2 \sim 10^{-8}$. Interestingly, in the mini-split SUSY scenario with gaugino masses, a 1-loop factor below the scalar masses~\cite{Arvanitaki:2012ps,ArkaniHamed:2012gw} is compatible with the experimental constraints on the SUSY flavor clockwork. The gauginos might be collider accessible.

\begin{acknowledgments}
The research of W.A. is supported by the U.S. Department of Energy grant number DE-SC0010107.
\end{acknowledgments}

\begin{appendix}
\section{Runnning of Gauge Couplings} \label{appendix}

The presence of the clockwork gear systems significantly alters the running of the gauge couplings above the clockwork scale. In fact, due to the large number of additional matter fields in the clockwork setup, asymptotic freedom is lost and the gauge couplings develop Landau poles above the clockwork scale. For example, the 1-loop renormalization group equation of the strong gauge coupling in the presence of the quark gear systems reads
\begin{equation}
    \frac{d \alpha_s}{d\log\mu} = 2 \left(\beta_0 + \sum_i \big( 2 N_q^i + N_u^i + N_d^i \big) \right)  \frac{\alpha_s^2}{4\pi} ~,
\end{equation}
with the well known $SU(3)_c$ 1-loop beta function coefficient of the MSSM $\beta_0 = -3$, and the sum runs over the three generations $i = 1, \dots ,3$. Using the gear numbers from~\eqref{eq:SUSYFlavorCWGearNumbers}, we find a large and positive beta function coefficient
\begin{equation}
\beta_0 + \sum_i \big( 2 N_q^i + N_u^i + N_d^i \big) = 24 -3n ~,
\end{equation}
for $n \in \{0,1,2,3\}$, determined by the magnitude of $\tan \beta$.
The corresponding location of the Landau pole,
\begin{equation}
    \Lambda_\text{LP} = \Lambda_\text{CW} \exp\left[ \frac{2\pi}{(24 - 3n) \alpha_s(\Lambda_\text{CW})} \right] ~,
\end{equation}
is approximately 3-4 orders of magnitude above the clockwork scale and depends weakly on the choice of $n$ or the value $\tan\beta$. Requiring that the strong gauge coupling does not develop a Landau pole below the Planck scale, $\Lambda_\text{LP} < M_\text{Pl} \sim 10^{19}$\,GeV, places a lower bound on the clockwork scale of around $\Lambda_\text{CW} \gtrsim 10^{16}$\,GeV. 

A qualitatively similar behavior is expected for the $SU(2)_L$ and $U(1)_Y$ gauge couplings. However, their running depends also on the details of the clockwork systems of the charged leptons and neutrinos. A quantitative discussion is therefore beyond the scope of this work. 

\end{appendix}

\bibliography{main}

\end{document}